\documentclass{article} 
\PassOptionsToPackage{dvipsnames}{xcolor}
\usepackage{iclr2022_conference,times}

\usepackage{amsmath,amsfonts,bm}

\def\eqref#1{equation~\ref{#1}}

\def\1{\bm{1}}

\DeclareMathAlphabet{\mathsfit}{\encodingdefault}{\sfdefault}{m}{sl}
\SetMathAlphabet{\mathsfit}{bold}{\encodingdefault}{\sfdefault}{bx}{n}

\usepackage{hyperref}
\usepackage{xurl}
\usepackage{booktabs}       
\usepackage{amsfonts}       
\usepackage{nicefrac}       
\usepackage{microtype}      
\usepackage{xcolor}         
\usepackage{amsmath}
\usepackage{multirow}
\usepackage{graphicx}
\usepackage{caption}
\usepackage{subcaption}
\usepackage{array}
\usepackage{wrapfig}
\usepackage{bbm}
\usepackage{inconsolata}
\usepackage{makecell}
\usepackage{comment}
\usepackage{fancyvrb,booktabs}
\usepackage{paralist}
\usepackage{tikz}
\usepackage{numprint}
\usepackage{multirow}
\usepackage{textcomp}
\usepackage{array}
\newcolumntype{H}{>{\setbox0=\hbox\bgroup}c<{\egroup}@{}}

\title{A Systematic Evaluation of Large Language Models of Code}

\author{Frank F. Xu, Uri Alon, Graham Neubig, Vincent J. Hellendoorn  \\
School of Computer Science\\
Carnegie Mellon University\\
\texttt{\{fangzhex,ualon,gneubig\}@cs.cmu.edu, vhellendoorn@cmu.edu} \\
}

\iclrfinalcopy 
\begin{document}

\maketitle

\begin{abstract}
    Large language models (LMs) of code have recently shown tremendous promise in completing code and synthesizing code from natural language descriptions.
    However, the current state-of-the-art code LMs (e.g., Codex \citep{chen2021evaluating}) are not publicly available, leaving many questions about their model and data design decisions.
    We aim to fill in some of these blanks through a systematic evaluation of  the largest existing models: Codex, GPT-J, GPT-Neo, GPT-NeoX-20B, and CodeParrot, across various programming languages.
    Although Codex itself is not open-source, we find that existing open-source models do achieve close results in some programming languages, although targeted mainly for natural language modeling.
    We further identify an important missing piece in the form of a large open-source model trained exclusively on a multi-lingual corpus of code. 
    We release a new model, PolyCoder, with 2.7B parameters based on the GPT-2 architecture, that was trained on 249GB of code across 12 programming languages on a single machine.
    In the C programming language, \emph{PolyCoder outperforms all models including Codex}.
    Our trained models are open-source and publicly available
    at \url{https://github.com/VHellendoorn/Code-LMs}, which enables future research and application in this area.
    \end{abstract}

\section{Introduction}


Language models (LMs) assign probabilities to sequences of tokens, and are widely applied to natural language text~\citep{bengio2003neural,baevski2018adaptive,brown2020language}.
Recently, LMs have shown impressive performance in modeling also source code, written in programming languages~\citep{hindle2016naturalness,hellendoorn2017deep,alon2020structural,karampatsis2020big}.
These models excel at useful downstream tasks like code completion~\citep{raychev2014code} and synthesizing code from natural language descriptions~\citep{desai2016program}.
The current state-of-the-art large language models for code, such as  \citet{austin2021program}, have shown significant progress for AI-based programming assistance. 
Most notably, one of the largest of these models, Codex~\citep{chen2021evaluating} has been deployed in the real-world production tool GitHub Copilot\footnote{\url{https://copilot.github.com/}}, as an in-IDE developer assistant that automatically generates code based on the user's context.

Despite the great success of large language models of code,  \emph{the strongest models are not publicly available}. This prevents the application of these models outside of well-resourced companies and limits research in this field for low-resourced organizations.
For example, Codex provides non-free access to the model's \emph{output} through black-box API calls,\footnote{\url{https://openai.com/blog/openai-codex/}} but the model's weights and training data are unavailable.
This prevents researchers from fine-tuning and adapting this model to domains and tasks other than code completion. 
The lack of access to the model's internals also prevents the research community from studying other key aspects of these models, such as interpretability, distillation of the model for more efficient deployment, and incorporating additional components such as retrieval.

Several medium to large-sized pre-trained language models are publicly available, such as GPT-Neo~\citep{gpt-neo}, GPT-J~\citep{gpt-j} and GPT-NeoX~\citep{gpt-neox-20b}.
Despite being trained on a mixture of a wide variety of text including news articles, online forums, and just a modest selection of (GitHub) software repositories~\citep{gao2020pile}, these language models can be used to generate source code with a reasonable performance~\cite{chen2021evaluating}.
In addition, there are a few open-source language models that are trained solely on source code. 
For example, CodeParrot~\citep{tunstall2022natural} was trained on 180 GB of Python code.

Given the variety of model sizes and training schemes involved in these models and lack of comparisons between these, the impact of many modeling and training design decisions remains unclear.
For instance, we do not know the precise selection of data on which Codex and other private models were trained; however, we do know that some public models (e.g., GPT-J) were trained on a mix of natural language and code in multiple programming languages, while other models (e.g., CodeParrot) were trained solely on code in one particular programming language.
Multilingual models potentially provide better generalization, because different programming languages share similar keywords and properties, as shown by the success of \emph{multilingual} models for natural language~\citep{conneau2019cross} and for code~\citep{zugner2021languageagnostic}.
This may hint that \emph{multilingual} LMs can \emph{generalize} across languages, outperform monolingual models and be useful for modeling low-resource programming languages, but this is yet to be verified empirically.

In this paper, we present a systematic evaluation of existing models of code -- Codex, GPT-J, GPT-Neo, GPT-NeoX, and CodeParrot -- across various programming languages.
We aim to shed more light on the landscape of code modeling design decisions by comparing and contrasting these models, as well as providing a key missing link: thus far, no large open-source language model was trained exclusively on code from \emph{multiple programming languages}. We provide three such models, ranging from 160M to 2.7B parameters, which we release under the umbrella name ``PolyCoder''.
First, we perform an extensive comparison of the training and evaluation settings between PolyCoder, open-source models, and Codex.
Second, we evaluate the models on the HumanEval benchmark~\citep{chen2021evaluating} and compare how do models of different sizes and training steps scale, and how different temperatures affect the generation quality.
Finally, since HumanEval only evaluates the natural language to Python synthesis, we curate an unseen evaluation dataset%
\footnote{The exact training set that Codex was trained on is unknown.}
in each of the 12 languages, to evaluate the perplexity of different models. 
We find that although Codex is allegedly focused on Python (\cite{chen2021evaluating} \textsection 3.1), Codex performs surprisingly well in other programming languages too, and even better than GPT-J and GPT-NeoX that were trained on the Pile \citep{gao2020pile}. Nonetheless, in the C programming language, \emph{our PolyCoder model achieves a lower perplexity than all these models, including Codex}.

Although most current models perform worse than Codex, we hope that this systematic study helps future research in this area to design more efficient and effective models. 
More importantly, through this systematic evaluation of different models, we encourage the community to study and release medium-large scale language models for code, in response to the concerns expressed by ~\citet{10.1145/3501261}: 

\makebox[\textwidth][c]{
\begin{minipage}{0.8\textwidth}
\emph{[...] this exploding trend in cost to achieve the state of the art has left the ability to train and test such models limited to a select few large technology companies—and way beyond the resources of virtually all academic labs}. 
\end{minipage}
}

We believe that our efforts are a significant step towards democratization of large language models of code.

\section{Related Work}
\label{sec:related}

\begin{figure}[t]
    \centering
    \includegraphics[width=0.8\textwidth]{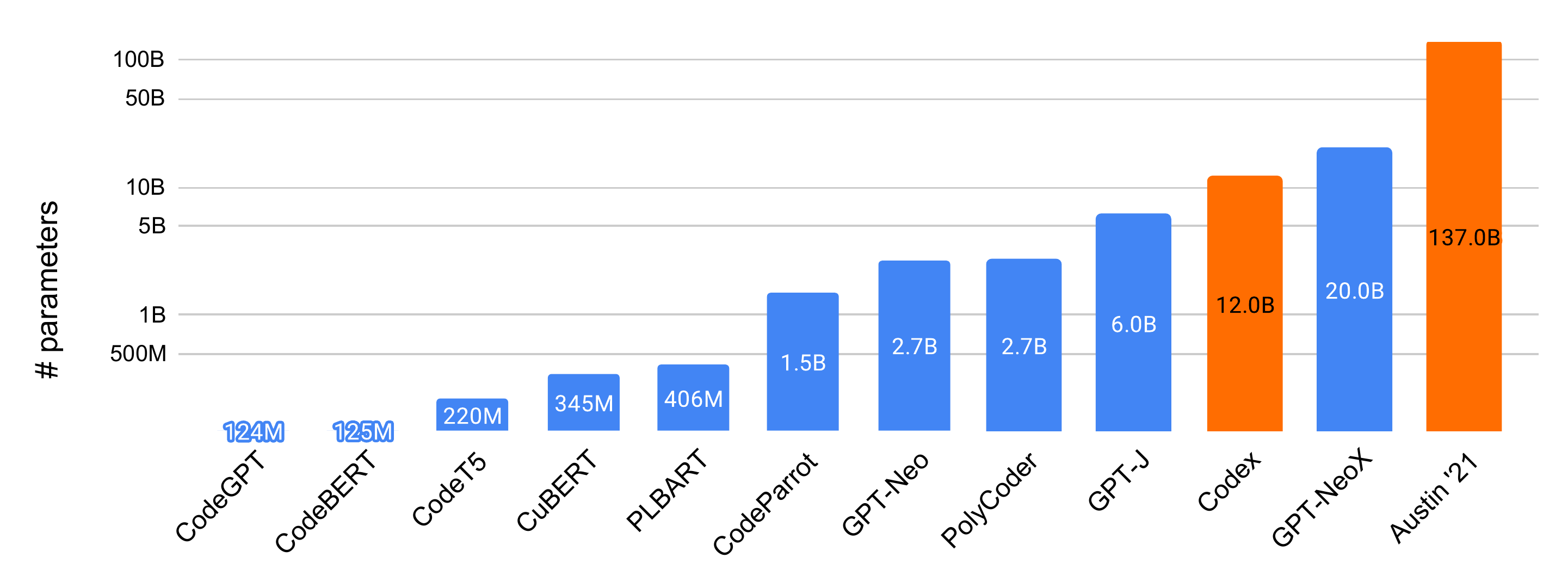}
	\caption{Existing language models of code, their sizes and availability (\colorbox{Cyan}{open source} vs. \colorbox{BurntOrange}{not open-source}).}
    \label{fig:related-barchart}
\end{figure}

\begin{figure}[t]
    \centering
    \includegraphics[width=1\textwidth]{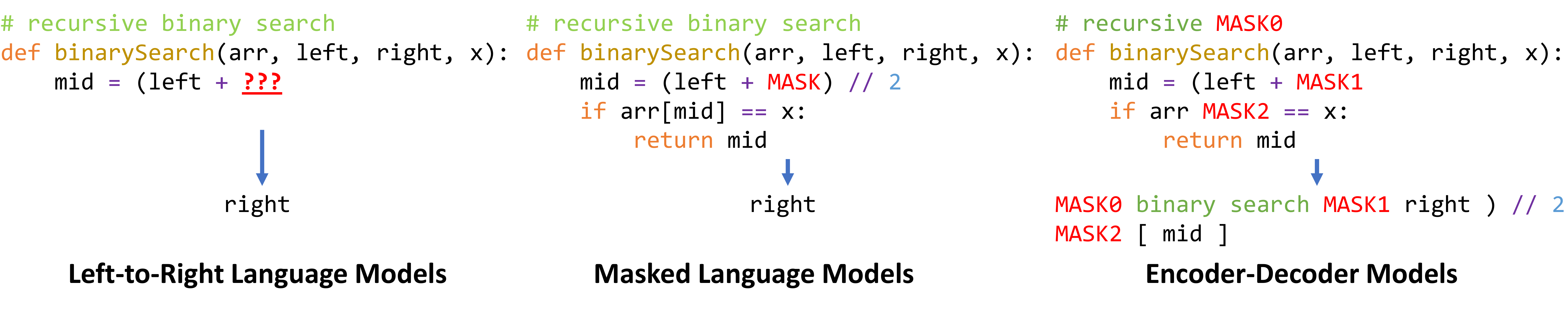}
	\caption{Three types of pretrained language models.}
    \label{fig:related-work}
\end{figure}

At the core of code modeling lies ongoing work on pretraining of language models (LMs).
Large-scale pretraining of LMs has had an astounding impact on natural language processing in recent years~\citep{han2021pre}.
Figure~\ref{fig:related-barchart} provides an overview of how different models compare in size and availability.

\subsection{Pretraining Methods}
We discuss three popular pretraining methods used in code language modeling.
An illustration of these methods are shown in Figure~\ref{fig:related-work}.

\paragraph{Left-to-Right Language Models} (Figure~\ref{fig:related-work}, left)
Auto-regressive, Left-to-right LMs, 
predict the probability of a token given the previous tokens.
In code modeling, CodeGPT (124M)~\citep{lu2021codexglue}, CodeParrot (1.5B)~\citep{tunstall2022natural}, GPT-Neo (2.7B)~\citep{gpt-neo}, GPT-J (6B)~\citep{gpt-j}, Codex (12B)~\citep{chen2021evaluating}, GPT-NeoX (20B)~\citep{gpt-neox-20b}, and Google's (137B)~\citep{austin2021program} belong to this category.
The left-to-right nature of these models makes them highly useful for program generation tasks, such as code completion.
On the other hand, as code is usually not written in a single, left-to-write pass, it is not trivial to leverage context that appears ``after'' the location of the generation.
In this paper, we focus on this family of models and will discuss the existing models in more detail in the following sections.

\paragraph{Masked Language Models}
(Figure~\ref{fig:related-work}, middle)
While auto-regressive language models are powerful for modeling the probability of sequences, their unidirectional nature makes them less suitable for producing effective whole-sequence representations for downstream tasks such as classification.
One popular bidirectional objective function used widely in representation learning is masked language modeling~\citep{devlin2018bert}, where the aim is to predict masked text pieces based on surrounding context.
CodeBERT (125M)~\citep{feng2020codebert} and CuBERT (345M)~\citep{kanade2020learning} are examples of such models in code.
In programming contexts, these methods provide useful representations of a sequence of code for downstream tasks such as code classification, clone detection, and defect detection.

\paragraph{Encoder-decoder Models}
(Figure~\ref{fig:related-work}, right)
An encoder-decoder model first uses an encoder to encode an input sequence, and then uses a left-to-right LM to decode an output sequence conditioned on the input sequence.
Popular pretraining objectives include masked span prediction~\citep{raffel2019exploring} where the input sequence is randomly masked with multiple masks and the output sequence are the masked contents in order, and denoising sequence reconstruction~\citep{lewis2019bart} where the input is a corrupted sequence and the output is the original sequence.
These pretrained models are useful in many sequence-to-sequence tasks~\citep{raffel2019exploring}.
In code, CodeT5 (220M)~\citep{wang2021codet5}, and PLBART (406M)~\citep{ahmad-etal-2021-unified} use the two objectives mentioned above respectively, and performs well in conditional generation downstream tasks such as code commenting, or natural language to code generation.

\subsection{Pretraining Data}
Some models (e.g. CodeParrot and CodeT5) are trained on  GitHub code only, with corpora extracted using either Google BigQuery's GitHub dataset~\footnote{\url{https://cloud.google.com/blog/topics/public-datasets/github-on-bigquery-analyze-all-the-open-source-code}}, or CodeSearchNet~\citep{husain2019codesearchnet}.
Others (e.g., GPT-Neo and GPT-J) are trained on ``the Pile"~\citep{gao2020pile}, a large corpus containing a blend of natural language texts and code from various domains, including Stack Exchange dumps, software documentations, and popular ($>$100 stars) GitHub repositories. 
The datasets on which other proprietary models (Codex, Google's) were trained on are unknown.
One goal of our study is to try to shed light on what corpora might be the most useful for pretraining models of code.

\section{Evaluation Settings}
We evaluate all models using both extrinsic and intrinsic benchmarks, as described below.

\paragraph{Extrinsic Evaluation}
One of the most popular downstream tasks for code modeling is code generation given a natural language description. 
Following \cite{chen2021evaluating}, we evaluate all  models on the HumanEval dataset.
The dataset contains 164 prompts with descriptions in the form of code comments and function definitions, including argument names and function names, and test cases to judge whether the generated code is correct.
To generate code given a prompt, we use the same sampling strategy as \citet{chen2021evaluating}, using softmax with a temperature parameter $\texttt{softmax}(x/T)$. 
We evaluate using a wide range of temperatures $T=[0.2,0.4,0.6,0.8]$ to control for the confidence of the model's predictions.
Similarly to Codex, we use nucleus sampling~\citep{holtzman2019curious} with top-$p = 0.95$.
We sample tokens from the model until we encounter one of the following
stop sequences that indicate the end of a method:\footnote{The absence of whitespace, which is significant in Python, signals an exit from the method body.} `\texttt{\textbackslash nclass}', `\texttt{\textbackslash ndef}', `\texttt{\textbackslash n\#}', `\texttt{\textbackslash nif}', or `\texttt{\textbackslash nprint}'. 
We randomly sample 100 examples per prompt in the evaluation dataset.

\paragraph{Intrinsic Evaluation}
To evaluate the intrinsic performance of different models, we compute the perplexity for each language on an unseen set of GitHub repositories. 
To prevent training-to-test data leakage for models such as GPT-Neo and GPT-J, we remove repositories in our evaluation dataset that appeared in the GitHub portion of the Pile training dataset~\footnote{\url{https://github.com/EleutherAI/github-downloader}}.
To evaluate Codex, we use OpenAI's API~\footnote{\url{https://beta.openai.com/docs/engines/codex-series-private-beta}}, choosing
the \texttt{code-davinci-001} engine.
We note that the data that this model was trained on is \emph{unknown}, so we cannot prevent data leakage from the training to the test set for Codex.
We sampled 100 random files for each of the 12 programming languages in our evaluation dataset.
To make perplexity comparable across different tokenization methods used in different models, we use Pygments~\footnote{\url{https://pygments.org/docs/lexers/}} 
to equally normalize the log-likelihood sum of each model, when computing perplexity.\footnote{Every model uses its original tokenizer for predicting the next token. We use the shared tokenizer only for computing the perplexity given the log-likelihood sum.}

\section{Compared Models}
\subsection{Existing Models}
As discussed in Section~\ref{sec:related}, we mainly focus on auto-regressive left-to-right pretrained language models, most suitable for code completion tasks.

We evaluate Codex, as it is currently deployed in real-world and has impressive performance in code completion~\citep{chen2021evaluating}.
Codex uses the GPT-3 language model~\citep{brown2020language} as its underlying model architecture.
Codex was trained on a dataset spanning 179GB (after deduplication) covering over 54 million public Python repositories obtained from GitHub on May 2020.
As reflected in its impressive results in other programming languages than Python, we suspect that Codex was also trained on large corpora of additional programming languages.
The model available for querying through a non-free API.

As for open-source models, we compare GPT-Neo, GPT-J and GPT-NeoX,
the largest variants having 2.7, 6 and 20 billion parameters, respectively.
GPT-NeoX is the largest open-source pretrained language models available.
These models are trained on the Pile dataset, so they are a good representatives of models that were trained on both natural language texts from various domains and source code from GitHub.
We also compare CodeParrot with at most 1.5 billion parameters, a model that was only trained on Python code from GitHub.
CodeParrot follows the process used in~\cite{chen2021evaluating} that obtained over 20M files Python files from Google BigQuery Github database, resulting in a 180GB dataset, which is comparable to Codex's \emph{Python} training data, but the model itself is much smaller.

There was no large open-source language model trained almost exclusively on code from multiple programming languages. 
To fill this gap, we train a 2.7 billion model, PolyCoder, on a mixture of repositories from GitHub in 12 different programming languages.

\begin{table}[ht]
    \centering
    \small
    \begin{tabular}{lrrrrH}
    \toprule
        Language & Repositories & Files & Size Before Filtering & Size After Filtering& Pile Size\\  \midrule
        C & 10,749 & 3,037,112 & 221G & 55G & 3.9G\\ 
        C\# & 9,511  & 2,514,494 & 30G & 21G & 3.0G\\ 
        C++ & 13,726  & 4,289,506 &115G & 52G& 4.7G \\ 
        Go & 12,371  & 1,416,789 &70G & 15G & 4.2G\\ 
        Java & 15,044  & 5,120,129 & 60G & 41G & 8.3G\\ 
        JavaScript & 25,144  & 1,774,174 & 66G & 22G & 17.9G\\ 
        PHP & 9,960  & 1,714,058 & 21G & 13G & 3.8G\\ 
        Python & 25,446  & 1,550,208& 24G & 16G & 12.9G\\ 
        Ruby & 5,826  & 674,343 & 5.0G & 4.1G & 3.1G\\ 
        Rust & 4,991  & 304,842 & 5.2G & 3.5G & 1.1G\\ 
        Scala & 1,497  & 245,100 & 2.2G & 1.8G & 0.6G\\ 
        TypeScript & 12,830  & 1,441,926 & 12G & 9.2G & 2.6G \\ \midrule
        Total & 147,095	&	24,082,681 & 631.4G & 253.6G & 95.2G\\
        \bottomrule
    \end{tabular}
    \caption{Training corpus statistics.}
    \label{tab:datastat}
\end{table}

\subsection{PolyCoder's Data}
\label{sec:data}
\paragraph{Raw Code Corpus Collection}
GitHub is an excellent source for publicly available source code of various programming languages. 
We cloned the most popular repositories for 12 popular programming languages with at least 50 stars (stopping at about 25K per language to avoid a too heavy skew towards popular programming languages) from GitHub in October 2021. 
For each project, each file belonging to the majority-language of that project was extracted, yielding the initial training set.
This initial, unfiltered dataset spanned 631GB and 38.9M files.

\paragraph{Data Preprocessing}
The detailed data preprocessing strategy comparison with other models are analyzed in Table~\ref{tab:datapre}.
In general, we tried to follow Codex's design decisions, although there is a fair bit of ambiguity in the description of its data preprocessing.

\paragraph{Deduplication and Filtering}
Similarly to Codex and CodeParrot, very large ($>$1MB) and very short ($<$100 tokens) files were filtered out, reducing the size of the dataset by 33\%, from 631GB to 424GB.
This only reduced the total \emph{number} of files by 8\%, showing that a small number of files were responsible for a large part of the corpus.%
\footnote{Codex additionally mentions removing ``auto-generated" files, but the definition of this was not clear, so we omitted this step.}

\cite{allamanis2019adverse} has shown that code duplication that commonly manifests in datasets of code adversely effects language modeling of code. 
Therefore, we deduplicated files based on a hash of their content, which reduced the number of files by nearly 30\%, and the dataset size by additional 29\%, leaving 24.1M files and 254GB of data.

Overall, the filtering of very large and very short files plus deduplication, reduced the number of files by 38\%, and the dataset size by 61\%, roughly on par with the 70\% dataset size reduction reported by CodeParrot.
A key difference that remains is that other approaches use more fine-grained filtering strategies, such as limiting the maximum line length or average line length, filtering of probable auto-generated files, etc.
 For example, \citet{chen2021evaluating} have filtered only 11\% of their training data.

The dataset statistics are shown in Table~\ref{tab:datastat}, showcasing data sizes per language before and after filtering.
Our dataset contains less Python code (only 16G) than Codex or CodeParrot, and instead covers many different programming languages. 

\paragraph{Tokenizer}
We train a GPT-2 tokenizer (using BPE~\citep{sennrich2015neural}) on a random 5\% subset of all the pretraining data, containing all the languages.
Codex uses an existing trained GPT-3 tokenizer, with the addition of multi-whitespace tokens to reduce the sequence length after tokenization, as consecutive whitespaces are more common in code than in text.

\begin{table}[t]
    \small
    \centering
    \begin{tabular}{p{1.5cm}p{3cm}p{4cm}p{4cm}}
    \toprule
    &PolyCoder & CodeParrot & Codex\\
    \midrule
        Dedup & Exact & Exact & Unclear, mentions ``unique''  \\ \midrule
        Filtering & Files $>$ 1 MB, $<$ 100 tokens & Files $>$ 1MB, max line length $>$ 1000, mean line length $>$ 100, fraction of alphanumeric characters $<$ 0.25, containing the word "auto-generated" or similar in the first 5 lines & Files $>$ 1MB, max line length $>$ 1000, mean line length $>$ 100, auto-generated (details unclear), contained small percentage of alphanumeric characters (details unclear)  \\ \midrule
        Tokenization & Trained GPT-2 tokenizer on a random 5\% subset (all languages) & Trained GPT-2 tokenizer on train split & GPT-3 tokenizer, add multi-whitespace tokens to reduce redundant whitespace tokens \\ \bottomrule
    \end{tabular}
    \caption{Comparison of data preprocessing strategies of different models.}
    \label{tab:datapre}
\end{table}

\subsection{PolyCoder's Training}
\label{sec:model}

\begin{figure}
\centering
\begin{subfigure}{.5\textwidth}
  \centering
  \includegraphics[width=1\linewidth]{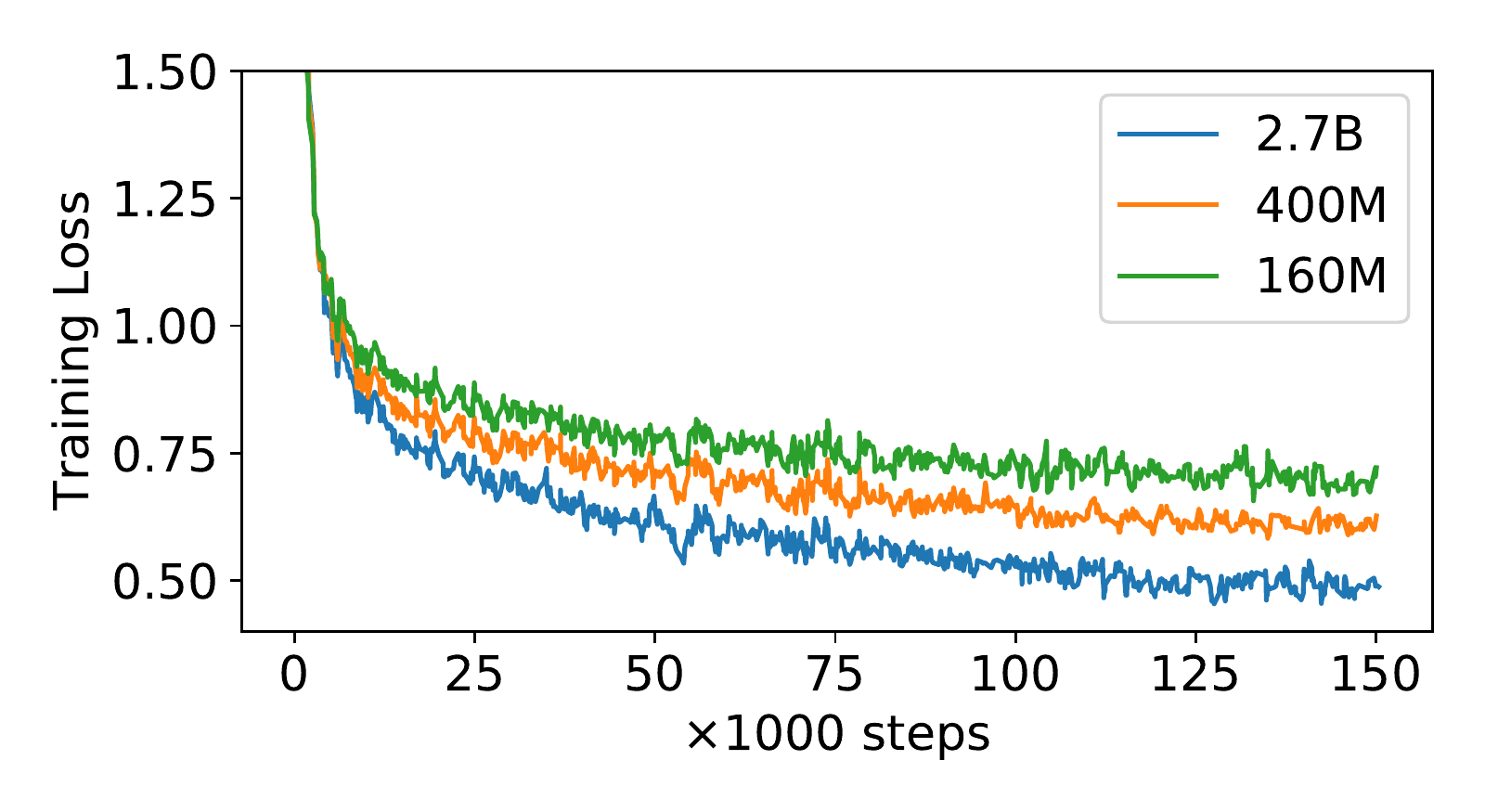}
  \caption{Training}
  \label{fig:train_loss}
\end{subfigure}%
\begin{subfigure}{.5\textwidth}
  \centering
  \includegraphics[width=1\linewidth]{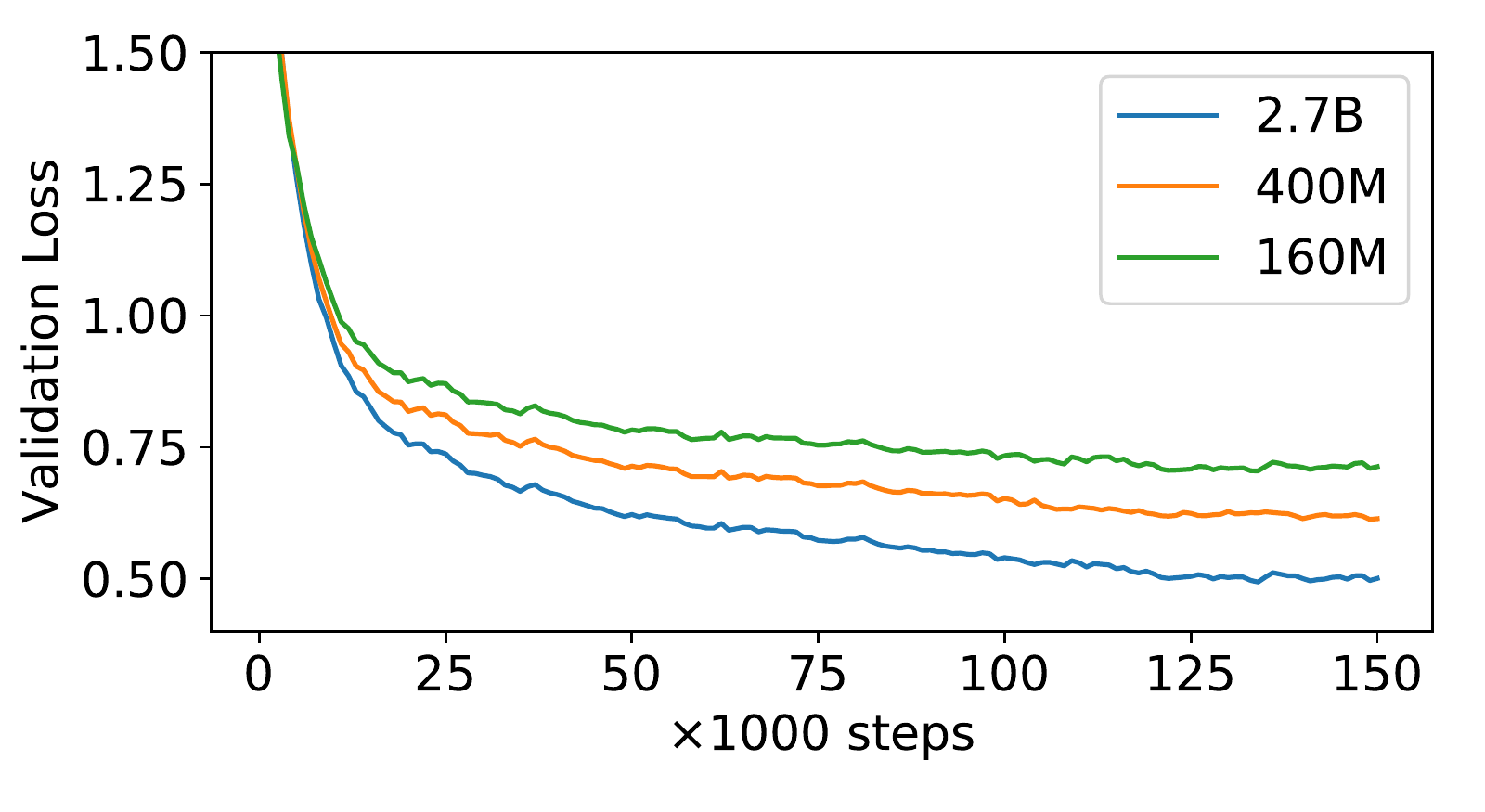}
  \caption{Validation}
  \label{fig:val_loss}
\end{subfigure}
\caption{Training and validation loss during the 150K step training process.}
\label{fig:training_curve}
\end{figure}

Considering our budget, 
we chose the GPT-2~\citep{radford2019language} as our model architecture.
To study the effect of scaling of model size, we train 3 different sized models, with 2.7 billion, 400 million and 160 million parameters, as the largest 2.7B model being on par with GPT-Neo for fair comparison.
The 2.7 billion model is a 32 layer, 2,560 dimensional Transformer model, with a max context window of 2048 tokens, trained with a batch size of 128 sequences (262K tokens). The model is trained for 150K steps.
The 400 million model is a 24 layer, 1,024 dimensional variant, and the 160 million model is a 12 layer, 768 dimensional variant, otherwise idem.
We use GPT-NeoX toolkit~\footnote{\url{https://github.com/EleutherAI/gpt-neox}} to train the model efficiently in parallel with 8 Nvidia RTX 8000 GPUs on a single machine.
The wall time used to train the largest 2.7B model is about 6 weeks. In its default configuration, this model should train for 320K steps, which was not feasible with our resources. Instead, we adjusted the learning rate decay to half this number and trained for up to 150K steps (near-convergence).
The training and validation loss curves for different sized models are shown in Figure~\ref{fig:training_curve}.
We see that even after training for 150K steps, the validation losses are still decreasing. This, combined with the shorter training schedule and faster learning rate decay, strongly signals that the models are still under-fitting and could benefit from longer training.

\begin{table}[t]
    \centering
    \small
    \begin{tabular}{p{3cm}p{3cm}p{3cm}p{3cm}}
    \toprule
    &PolyCoder (2.7B) & CodeParrot (1.5B) & Codex (12B)\\
    \midrule
        Model Initialization & From scratch & From scratch & Initialized from GPT-3 
  \\ 
           NL Knowledge& Learned from comments in the code & Learned from comments in the code & Natural language knowledge from GPT-3
  \\  \midrule
    Learning Rate& 1.6e-4 & 2.0e-4 & 1e-4
  \\ 
    Optimizer& AdamW & AdamW & AdamW
  \\ 
    Adam betas& 0.9, 0.999 & 0.9, 0.999 & 0.9, 0.95
  \\ 
    Adam eps& 1e-8 & 1e-8 & 1e-8
  \\ 
    Weight Decay& - & 0.1 & 0.1
  \\ 
    Warmup Steps& 1600 & 750 & 175
  \\ 
    Learning Rate Decay& Cosine & Cosine & Cosine
  \\  \midrule
    Batch Size (\#tokens)& 262K & 524K & 2M
  \\ 
    Training Steps& 150K steps, 39B tokens & 50K steps, 26B tokens & 100B tokens \\
    Context Window& 2048 & 1024 & 4096 \\
    \bottomrule
  \\ 
    \end{tabular}
     \caption{Comparison of design decisions and hyper-parameters in training different models of code.}
    \label{tab:hyperparam}
\end{table}

We compare the training setting and hyperparameters with CodeParrot and Codex in Table~\ref{tab:hyperparam}.
Due to high computational costs, we were unable to perform hyperparameter search.
Most hyperparameters are the same as those used in their respective GPT-2 model training~\footnote{\url{https://github.com/EleutherAI/gpt-neox/tree/main/configs}} to provide a good default with regards to the corresponding model size.
Some key differences include context window sizes to allow for more tokens as context, batch sizes and tokens trained, as well as model initialization with or without natural language knowledge.

\section{Results}
\label{sec:results}
\subsection{Extrinsic Evaluation}
\label{subsec:extrinsic}

\begin{table}[t]
    \centering
    \small
    \begin{tabular}{lrrrccc}
    \toprule
    Model & Pass@1 & Pass@10 & Pass@100 & Tokens Trained & Code Tokens & Python Tokens \\
    \midrule
    PolyCoder (160M) & 2.13\% & 3.35\% & 4.88\% &39B  &39B  &2.5B\\ 
    PolyCoder (400M) & 2.96\% & 5.29\% & 11.59\% &39B  &39B &2.5B \\ 
    PolyCoder (2.7B) & 5.59\% & 9.84\% & 17.68\% &39B  &39B &2.5B\\ 
    \midrule
    CodeParrot (110M)& 3.80\% & 6.57\% & 12.78\% & 26B  & 26B  & 26B \\
    CodeParrot (1.5B)& 3.58\% & 8.03\% & 14.96\% & 26B  & 26B  & 26B\\
    \midrule
    GPT-Neo (125M) & 0.75\% & 1.88\% & 2.97\% &300B & 22.8B & 3.1B \\
    GPT-Neo (1.3B) & 4.79\% & 7.47\% & 16.30\% &380B & 28.8B & 3.9B \\
    GPT-Neo (2.7B) & 6.41\% & 11.27\% &  21.37\% &420B & 31.9B & 4.3B\\
    GPT-J (6B) & 11.62\% & 15.74\% &  27.74\% &402B & 30.5B & 4.1B\\
    \midrule
    Codex (300M)	& 13.17\% & 20.37\% & 36.27\% &100B* &100B* &100B*\\
    Codex (2.5B)	& 21.36\% & 35.42\% & 59.50\% &100B* &100B* &100B*\\
    Codex (12B)	& 28.81\% & 46.81\% & 72.31\% &100B* &100B* &100B* \\
    \bottomrule
    \multicolumn{7}{r}{\footnotesize{*Codex is initialized with another pretrained model, GPT-3.}} \\
    \end{tabular}
    \caption{Results of different models on the HumanEval benchmark, and the number of different types of tokens seen during the training process.}
    \label{tab:humaneval_res}
\end{table}

The overall results are shown in Table~\ref{tab:humaneval_res}.\footnote{Due to the large model size of GPT-NeoX (20B) and limited computational budget, we did not include it in the HumanEval experiment.}
The numbers are obtained by sampling with different temperatures and picking the best value for each metric.
Among existing models, PolyCoder is worse than similarly sized GPT-Neo and the even smaller Codex 300M.
Overall, PolyCoder lies after Codex, GPT-Neo/J, while performing stronger than CodeParrot.
PolyCoder, which was trained only on code, falls behind a similar sized model (GPT-Neo 2.7B) trained on the Pile, a blend of natural language texts and code.  
Looking at the rightmost columns in Table~\ref{tab:humaneval_res} offers a potential explanation: in terms of total Python tokens seen during training, all models substantially exceed ours. 
This in partly because they use a higher proportion of Python code (we aimed to balance data volume across programming languages), and in part because of resource limitations, which lead to PolyCoder not observing its entire training data. 
In addition, the natural language blend in the training corpus may help code language modeling as well, especially with code-related texts such as Stack Exchange dumps being included.

Compared to GPT-Neo (2.7B), PolyCoder has seen fewer Python tokens, but more code tokens in other programming languages, hinting that transfer from other languages to Python helps to achieve a similar performance.
This suggests that future research could benefit from blending code in different programming languages, as well as natural language text. 

\begin{figure}
\centering
\begin{subfigure}{.33\textwidth}
  \centering
  \includegraphics[width=1\linewidth]{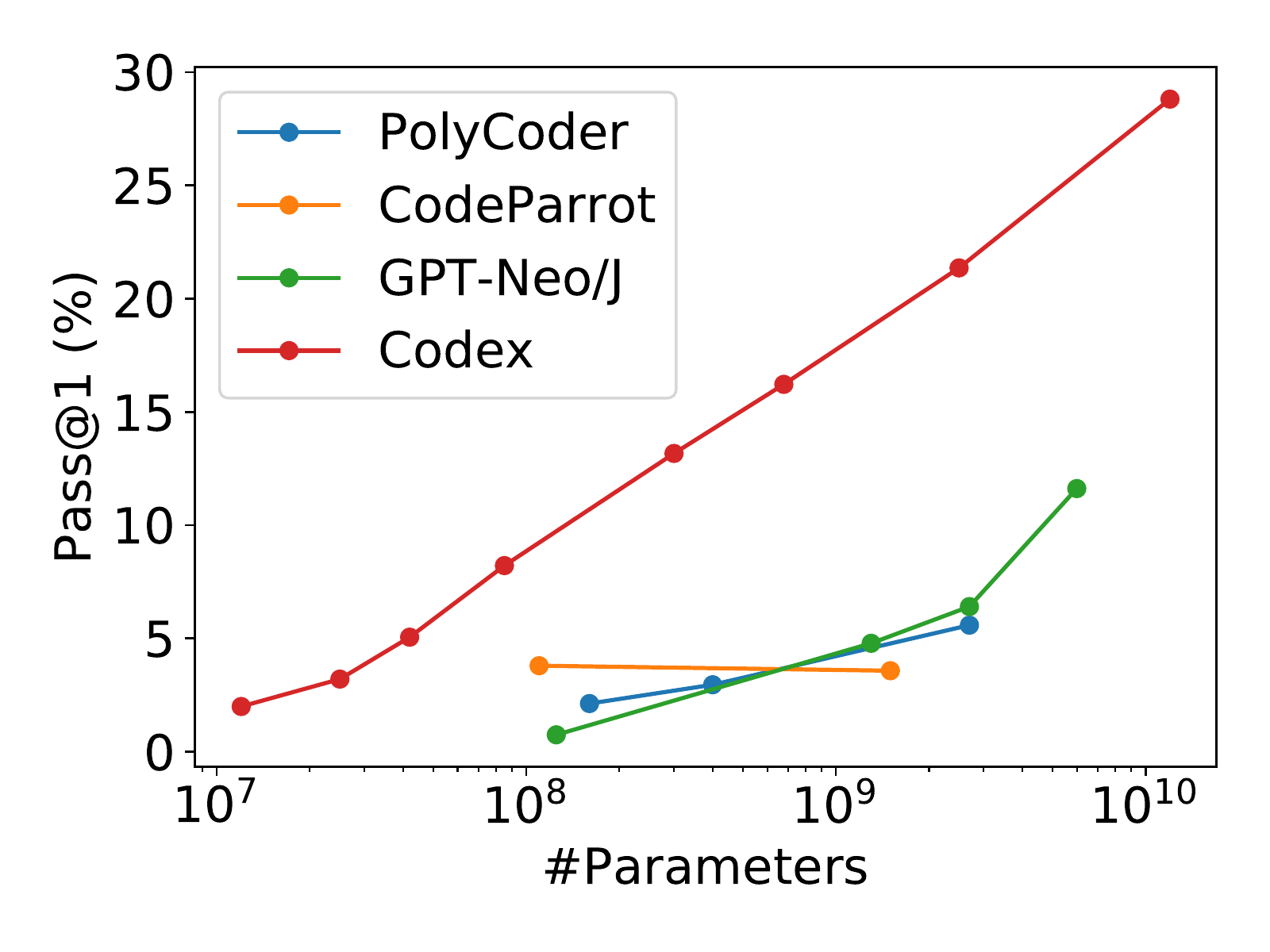}
  \caption{Pass@1}
  \label{fig:pass1_param}
\end{subfigure}%
\begin{subfigure}{.33\textwidth}
  \centering
  \includegraphics[width=1\linewidth]{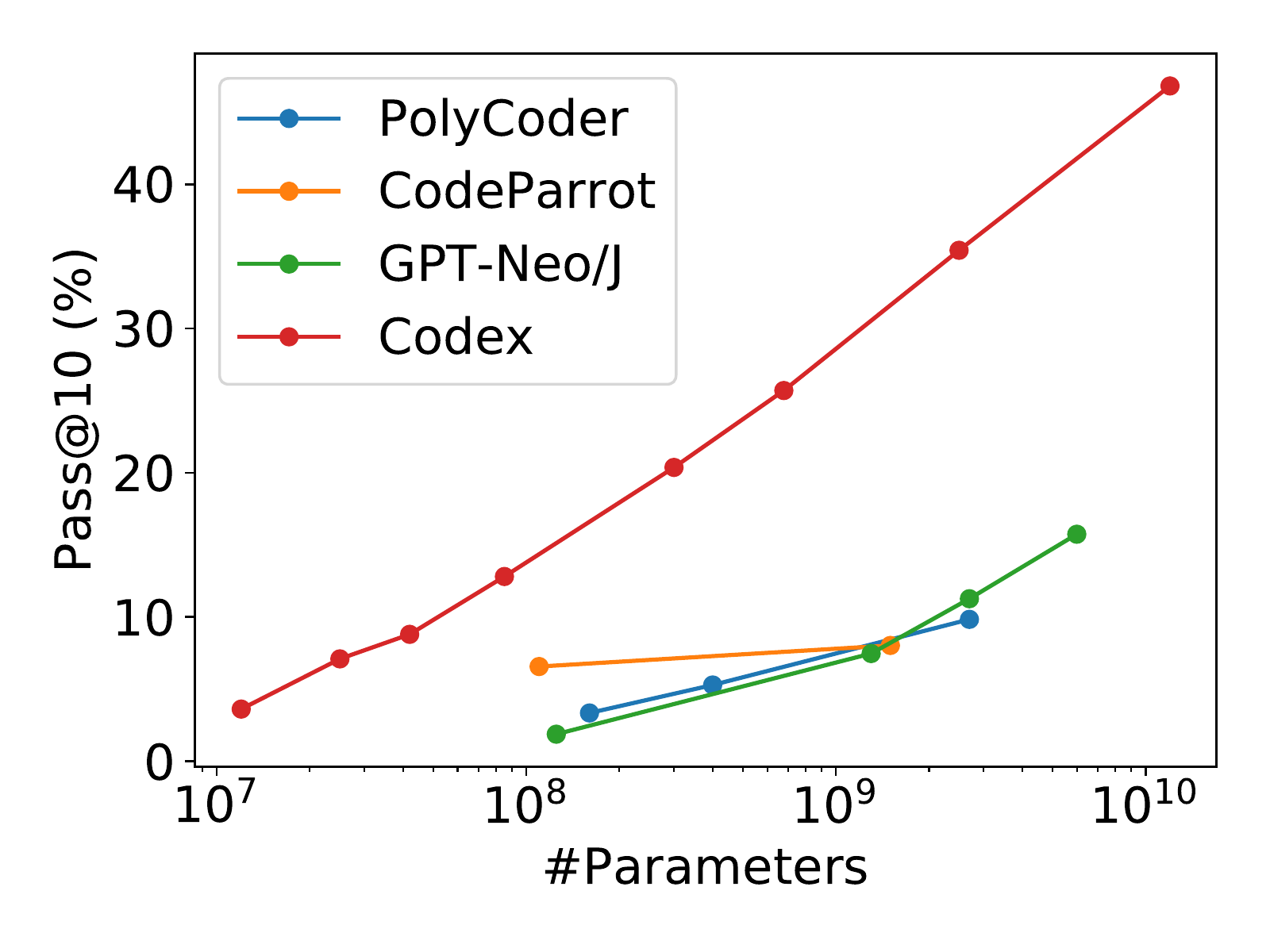}
  \caption{Pass@10}
  \label{fig:pass10_param}
\end{subfigure}
\begin{subfigure}{.33\textwidth}
  \centering
  \includegraphics[width=1\linewidth]{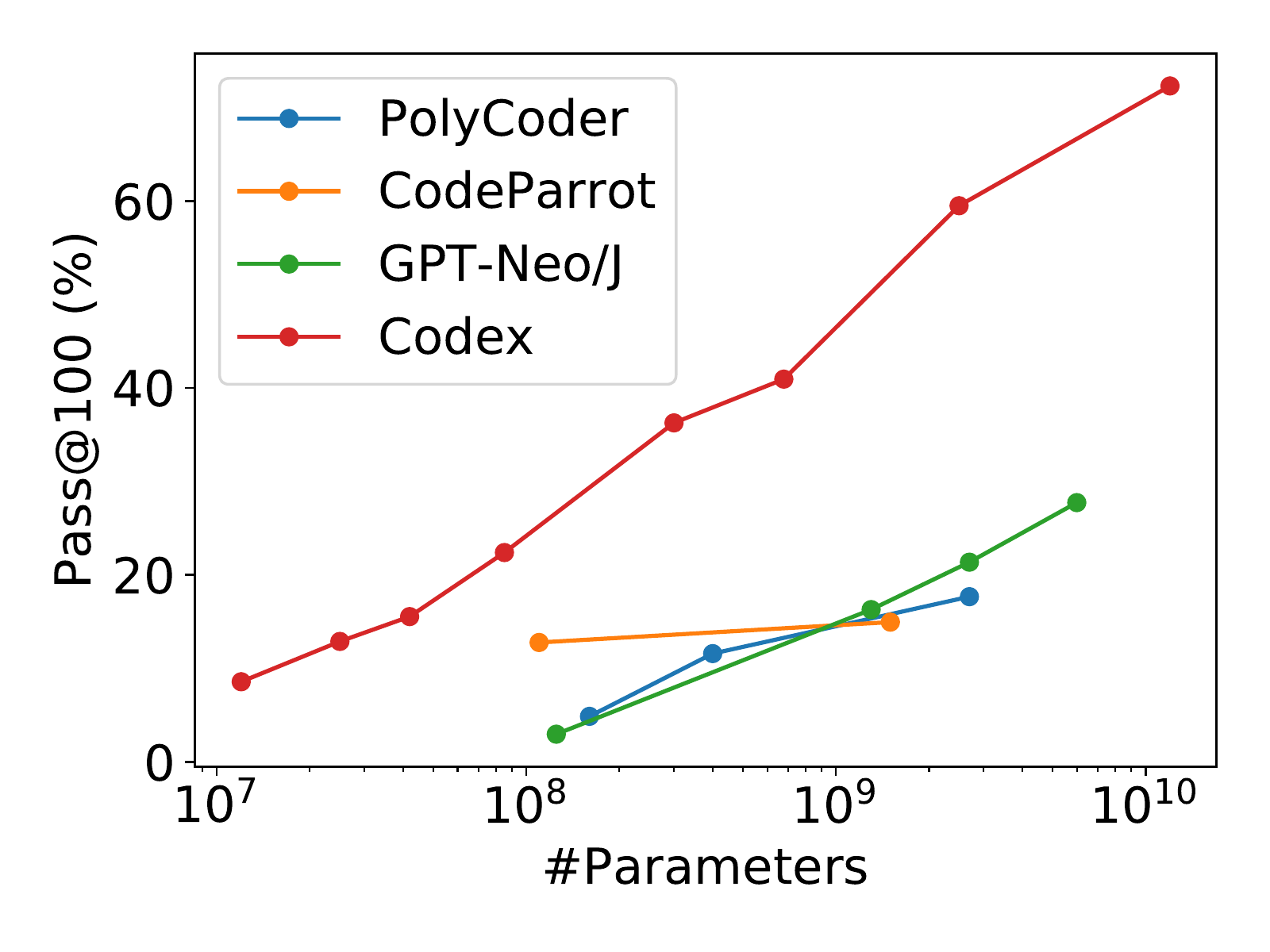}
  \caption{Pass@100}
  \label{fig:pass100_param}
\end{subfigure}
\caption{The scaling effect of HumanEval performance on different models.}
\label{fig:performance_param}
\end{figure}
\paragraph{Scaling Effect} To further understand the effect of the number of model parameters with respect to HumanEval code completion performance, we show the Pass@1, Pass@10 and Pass@100 percentage with respect the the model size in Figure~\ref{fig:performance_param}.
We can see that the performance of the Codex models  are significantly better than all the other open-source models across all numbers of parameters.
The performance on HumanEval benchmark increases linearly with the magnitude (log scale) of the number of parameters in the model.
Similar scaling effects could be found on PolyCoder and GPT-Neo/J models. 
Interestingly, the CodeParrot models that are trained only on Python  seem to have reached a saturating performance with respect to increasing number of parameters, where the training corpus being focused on Python may have some effect.
With higher number of parameters (2.7B), PolyCoder's performance is trending worse than that of GPT-Neo/J.
Comparing GPT-Neo/J that is trained on Pile dataset containing a blend of text, Stack Exchange dumps and GitHub data, with PolyCoder that are trained on only GitHub repositories of popular programming languages, we hypothesize that the added text, especially texts in technical and software engineering domains, may be crucial for the larger model to boost the performance.
We also compare the performance difference between the model trained after 100K steps versus the model after 150K steps in Appendix~\ref{app:trainlonger}, and find that training for longer helps the larger model more as it is still under-fitted.

\begin{wrapfigure}{r}{0.4\textwidth}
\centering
  \includegraphics[width=1\linewidth]{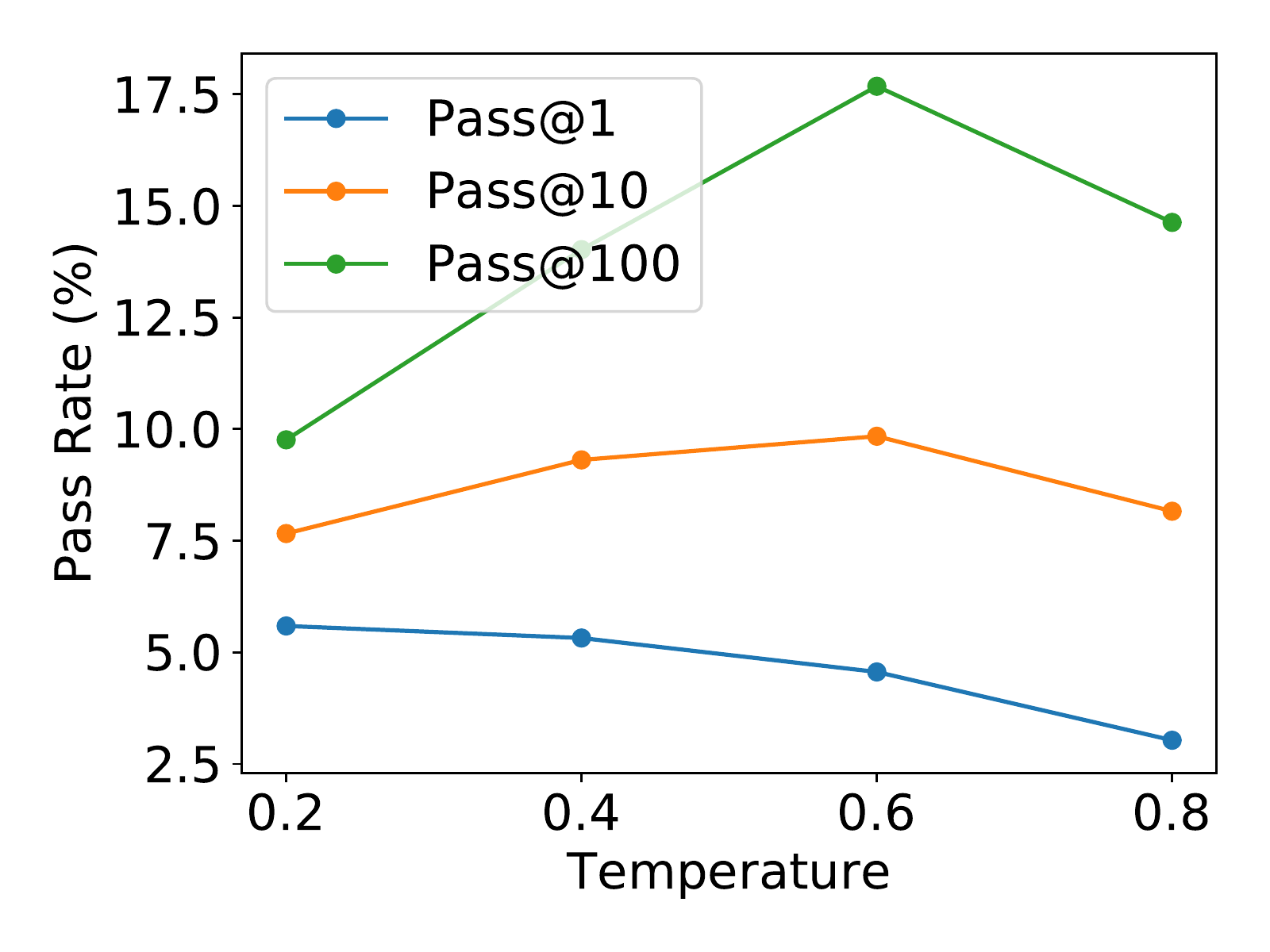}
\caption{HumanEval performance with different softmax temperatures during generation.}
\label{fig:temperature}
\vspace{-10mm}
\end{wrapfigure}

\paragraph{Temperature Effect}
All the above results are obtained by sampling the language model with different temperatures and picking the best value for each metric.
We are also interested in how different choices of temperature affects the final generation quality.
We summarize the results in Figure~\ref{fig:temperature}.
The general trend is for Pass@1, lower temperatures are better, and for Pass@100, a higher temperature will help, while for Pass@10 a temperature in the middle is better suited.
We hypothesize that this is because a higher temperature during generation makes the model less confident in its predictions and thus allow for more exploration and more diverse outputs, resulting in better accuracy at Pass@100.
Too high a temperature (0.8) is also hurtful if the model is capable enough.
On the contrary, a lower temperature makes the model output very confident in its prediction and thus will be better suited for generating very few correct examples, and thus the better performance for Pass@1.
In Appendix~\ref{app:tempsmaller} we repeat these experiments with the smaller models as well.
This suggests the importance of temperature and the need to tune it individually for different generation scenarios.

\begin{figure}[t]
    \centering
    \includegraphics[width=1\textwidth]{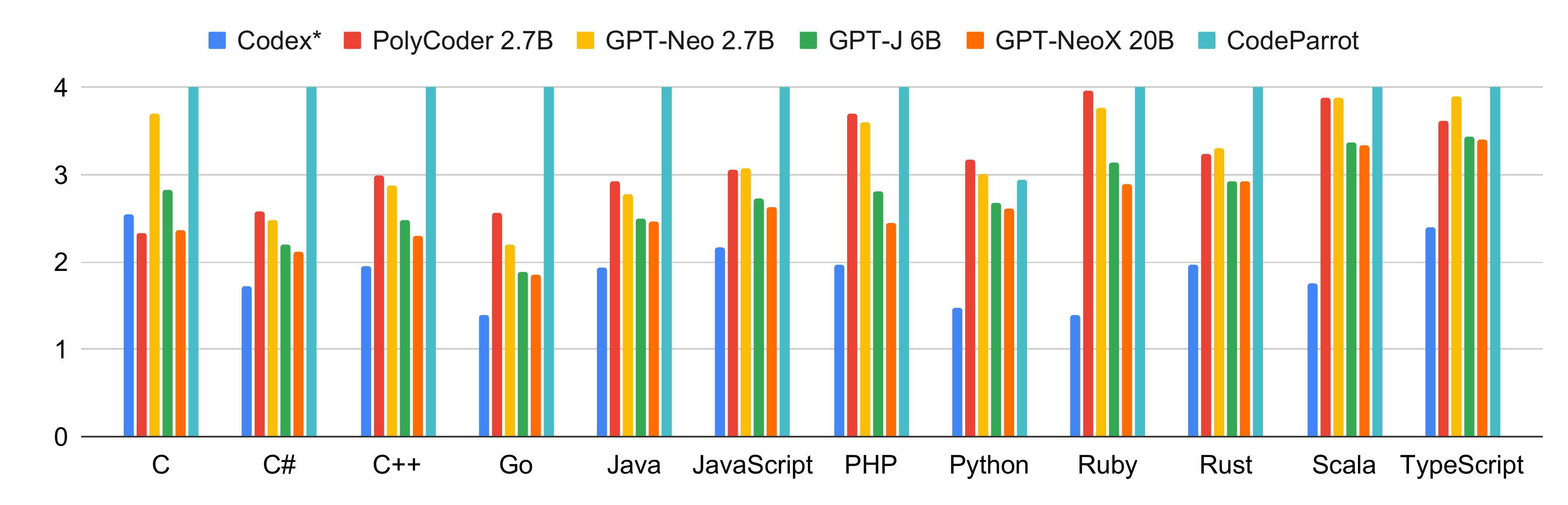}
    \footnotesize{* Since the exact training set of Codex is unknown, it may include files from these test sets} \\
\footnotesize{ rendering Codex's results overly-optimistic.}
	\caption{Perplexity comparison on our evaluation dataset of different models on different programming languages. Note that the y-axis is capped at 4; CodeParrot's entropy on all languages other than Python is much higher than shown here (see Table \ref{tab:perplexity}).}
    \label{fig:ppl-barchart}
\end{figure}

\subsection{Intrinsic Evaluation}
The perplexity results on the evaluation datasets are shown in Figure~\ref{fig:ppl-barchart}, with detailed numbers in Appendix~\ref{app:ppldetail}.
The plot caps the perplexity score to 4 as CodeParrot performs poorly in languages other than Python.
It is important to note that although Codex's perplexities are lower than other models in most languages, Codex might have been trained on the test sets, and its results are thus over-optimistic.

Notably, \emph{PolyCoder outperforms Codex and all other models in the C language}. Comparing the open-source models only, PolyCoder performs better than the similarly sized GPT-Neo 2.7B in C, JavaScript, Rust, Scala and TypeScript.


In the other 11 languages other than C, all other open-source models, including ours, are significantly worse (higher perplexity) than Codex.
We hypothesize that this is due to the fact that PolyCoder is trained on an imbalanced mixture of different languages, with C and C++ being closely related and the two most dominant in the entire training corpus (Section~\ref{sec:data}).
Thus, the larger volume in total (because of long files) makes C the most ``favored'' language by PolyCoder. 
The reason why PolyCoder does not outperform Codex in C++ is possibly due to the complexity of C++ language and Codex's significantly longer context window size (4096, compared to PolyCoder's 2048),
or because Codex is possibly trained on more C++ training data. 

With the same pretraining corpus, the gain from a 2.7B model (GPT-Neo) to a 6B model (GPT-J) is significant over all languages.
However, when increasing the model size further to 20B, the improvement varies across different languages.
For example, the performance on Go, Java, Rust, Scala, TypeScript do not increase significantly when the model size increases by 3 times.
This suggests that for some programming languages, and given the amounts of data, the capacity of GPT-J is sufficient.
Interestingly, these languages seem to coincide with languages where PolyCoder outperforms a similarly sized model trained on Pile.
This may hint that for the languages in which larger models do not provide additional gains, training the model only using code may be enough or slightly more helpful than training on both natural language and code.

We can see that comparing different models, perplexity trends for Python correlates well with the HumanEval benchmark performance of the extrinsic evaluation (Section  \ref{subsec:extrinsic}).
This suggests that perplexity is a useful and low-cost metric to estimate other, downstream, metrics.

\section{Conclusion}
In this paper, we perform a systematic evaluation of large language models for code.
The performance generally benefits from larger models and longer training time. 
We also believe that the better results of GPT-Neo over PolyCoder in some languages show that training on natural language text \emph{and} code can benefit the modeling of code.
To help future research in the area, we release PolyCoder, a large open-source language model for code, trained exclusively on code in 12 different programming languages. In the C programming language, \emph{PolyCoder achieves lower perplexity than all models including Codex}.

\bibliography{ref}
\bibliographystyle{iclr2022_conference}

\appendix
\section{Scaling Effect: Trained Longer}
\label{app:trainlonger}
\begin{figure}[h]
\centering
\begin{subfigure}{.33\textwidth}
  \centering
  \includegraphics[width=1\linewidth]{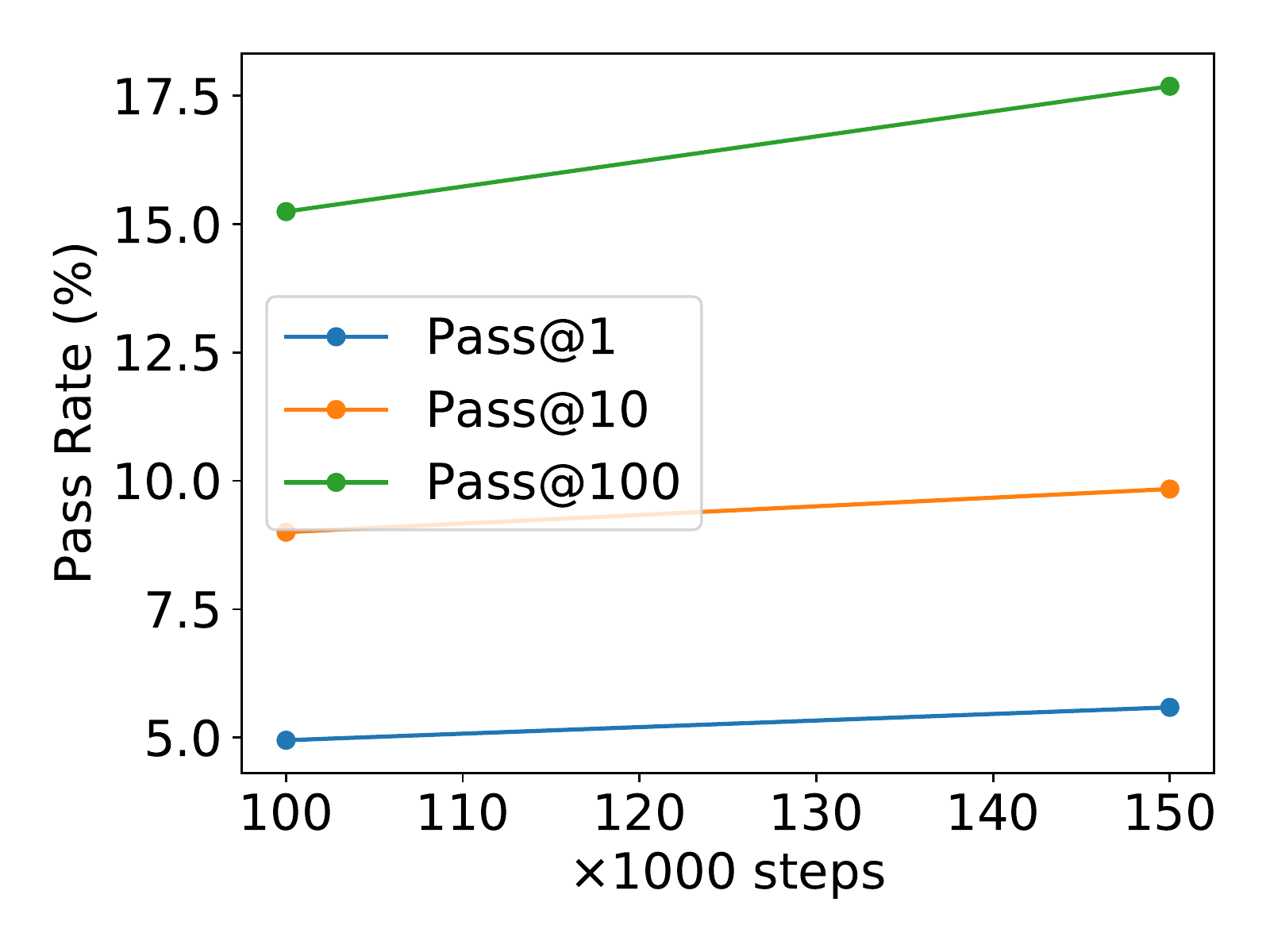}
  \caption{2.7B Model}
  \label{fig:2.7B_train_steps}
\end{subfigure}%
\begin{subfigure}{.33\textwidth}
  \centering
  \includegraphics[width=1\linewidth]{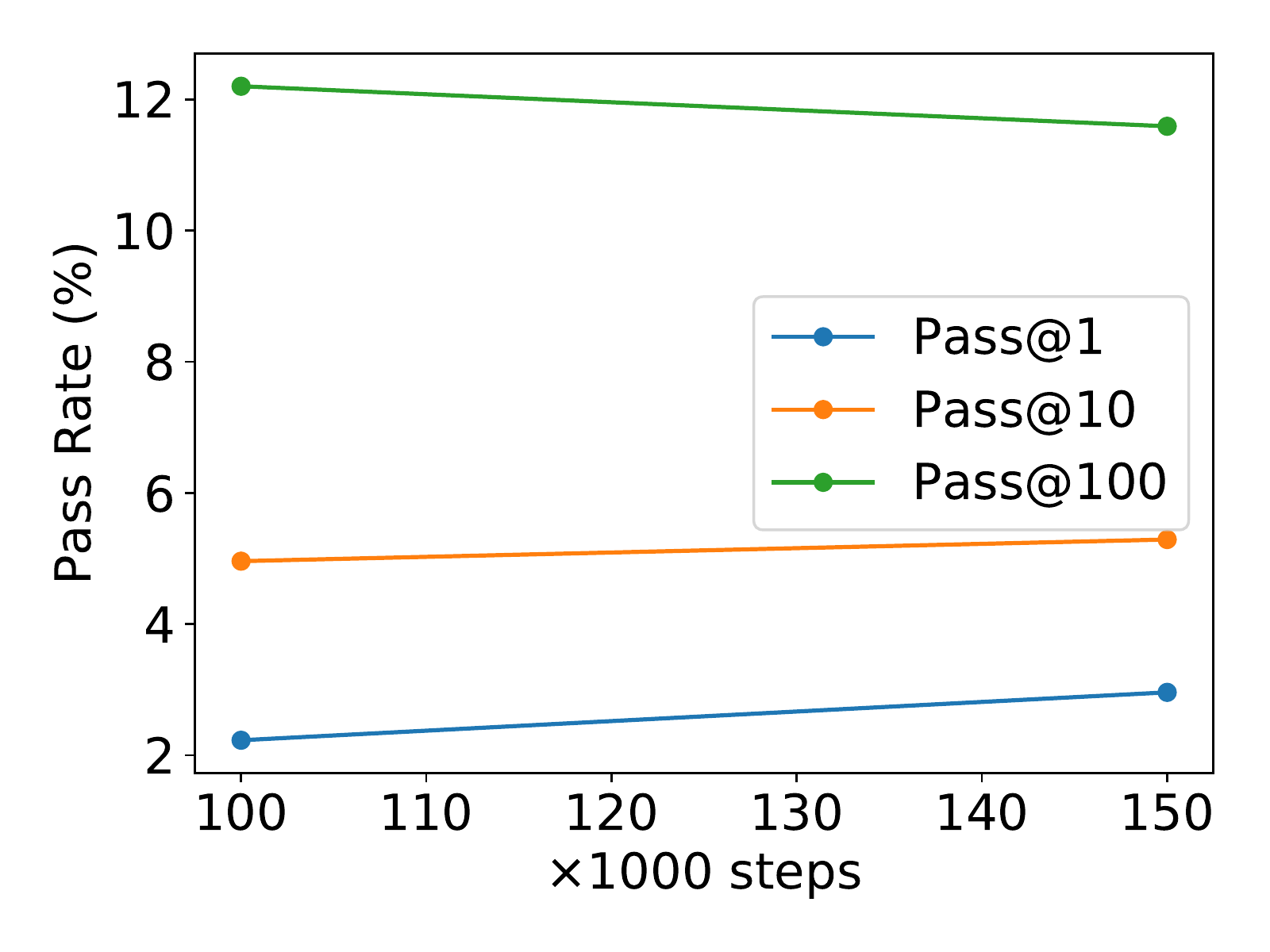}
  \caption{400M Model}
  \label{fig:400M_train_steps}
\end{subfigure}
\caption{HumanEval performance comparison after training the model for longer.}
\label{fig:steps_performance}
\end{figure}

We compare the performance difference between the model trained after 100K steps versus the model after 150K steps in Figure~\ref{fig:steps_performance}.
We can see that in the larger 2.7B model, by training the model longer till 150K steps, the performance increases uniformly, with Pass@100 increasing the most.
However, for a smaller model such as the 400M model, by training the model longer till 100K steps, the improvements are subdued and Pass@100 drops. 
This suggests that with the larger model, training for longer may provide additional boost in performance.
This echoes with the observation from the training curve (Figure~\ref{fig:training_curve}) as well. 

\section{Temperature Effect: Smaller Models}
\label{app:tempsmaller}

\begin{figure}[h]
\centering
\begin{subfigure}{.32\textwidth}
  \centering
  \includegraphics[width=1\linewidth]{figures/2.7B_temperature.pdf}
  \caption{2.7B Model}
  \label{fig:2.7B_temperature}
\end{subfigure}%
\begin{subfigure}{.32\textwidth}
  \centering
  \includegraphics[width=1\linewidth]{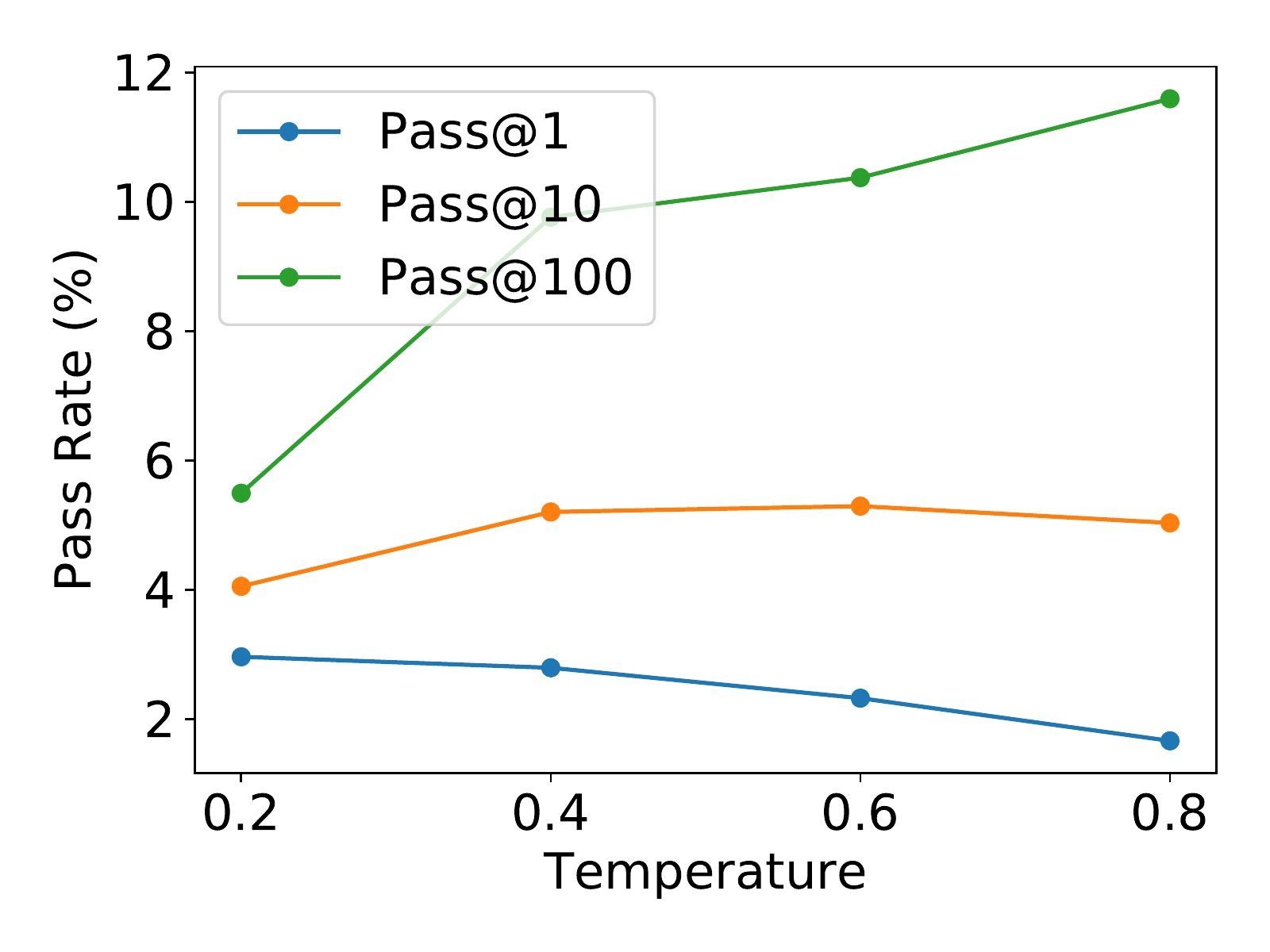}
  \caption{400M Model}
  \label{fig:400M_temperature}
\end{subfigure}
\begin{subfigure}{.32\textwidth}
  \centering
  \includegraphics[width=1\linewidth]{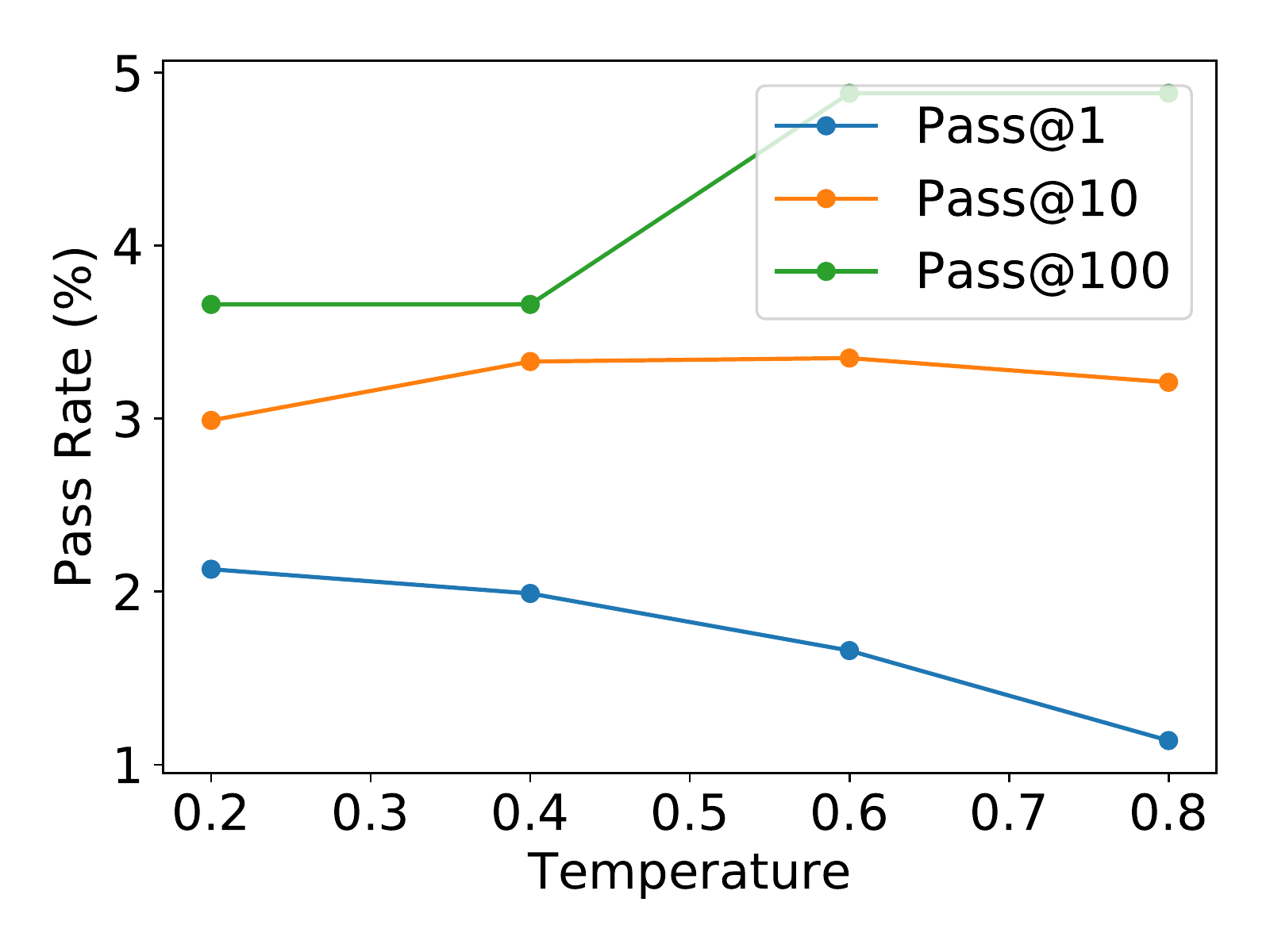}
  \caption{160M Model}
  \label{fig:160M_temperature}
\end{subfigure}
\caption{HumanEval performance using different softmax temperatures during generation.}
\label{fig:alltemperature}
\end{figure}

We show how temperature affects HumanEval performance on model of all three sizes in Figure~\ref{fig:alltemperature}. 
We find that for a larger model, e.g., the 2.7B model, a temperature as high as 0.8 is actually hurting the performance for Pass@100, suggesting that if the model is good enough, a very high temperature may cause the outputs to be too diverse, thus hurting the correctness.
This suggests the importance of temperature and the need to tune it individually for different model capacity and different generation scenarios. 

\section{Detailed Perplexity Results}
\label{app:ppldetail}
We show the detailed perplexity of different models on different languages in Table~\ref{tab:perplexity}.
The number of tokens shown in the table is obtained after tokenizing the code in each language using their respective lexers, by Pygments.
This number of tokens is used to normalize the perplexity scores to make them comparable across models.
Note that CodeParrot is only trained on Python data and thus performs poorly in other languages.

\begin{table}[t]
\centering
\small
\begin{tabular}{lrrrrrrr}
\toprule
Language & \multicolumn{1}{l}{\#tokens} & \multicolumn{1}{l}{Codex*} & \multicolumn{1}{l}{PolyCoder 2.7B} & \multicolumn{1}{l}{GPT-Neo 2.7B} & \multicolumn{1}{l}{GPT-J 6B} & \multicolumn{1}{l}{GPT-NeoX} & \multicolumn{1}{l}{CodeParrot} \\\midrule
C          & 55,333                       & 2.55                        & 2.33                        & 3.69                           & 2.82                       & 2.37                           & 19.23                        \\
C\#         & 67,306                       & 1.72                        & 2.58                        & 2.49                           & 2.20                       & 2.12                           & 7.16                        \\
C++        & 69,627                       & 1.95                        & 2.99                        & 2.87                           & 2.47                       & 2.32                           & 8.48                        \\
Go         & 79,947                       & 1.39                        & 2.57                         & 2.19                           & 1.89                       & 1.85                           & 10.00                        \\
Java       & 65,484                       & 1.94                        & 2.92                        & 2.78                           & 2.49                       & 2.47                            & 6.79                        \\
JavaScript & 54,620                       & 2.17                        & 3.06                       & 3.07                           & 2.73                       & 2.62                           & 9.23                        \\
PHP        & 45,682                       & 1.98                        & 3.70                        & 3.61                           & 2.81                       & 2.45                           & 19.91                        \\
Python     & 79,653                       & 1.47                        & 3.18                        & 3.00                           & 2.68                       & 2.61                           & 2.95                          \\
Ruby       & 46,537                       & 1.39                        & 3.96                       & 3.77                           & 3.13                       & 2.89                          & 14.26                        \\
Rust       & 107,717                      & 1.96                        & 3.24                        & 3.30                           & 2.92                       & 2.92                           & 8.68                        \\
Scala      & 65,756                       & 1.75                        & 3.87                       & 3.88                           & 3.37                      & 3.33                          & 12.91                        \\
TypeScript & 55,895                       & 2.40                        & 3.61                        & 3.90                            & 3.43                      & 3.41                           & 12.54          \\
\bottomrule
\multicolumn{8}{l}{\footnotesize{* Since the exact training set of Codex is unknown, it might have been trained on these test sets,}} \\
\multicolumn{8}{l}{\footnotesize{ and Codex's results are over-optimistic.}}
\end{tabular}
\caption{Perplexity of different models for different programming languages on our evaluation dataset.
}
\label{tab:perplexity}
\end{table}

\end{document}